\documentclass[fleqn,usenatbib]{mnras}

\usepackage{newtxtext,newtxmath}

\usepackage[T1]{fontenc}
\usepackage{ae,aecompl}

\usepackage{graphicx}	
\usepackage{amsmath}	

\newcommand{\dss} {$\delta$ Scuti stars}
\newcommand{\ds} {$\delta$ Scuti}

\newcommand{\rhom} {${\bar \rho}$}

\newcommand{\teff} {T$_{\mathrm{eff}}$}
\newcommand{\logg}{$\log g$}

\newcommand{\Dnu}{$\Delta\nu$}

\newcommand{\most} {\emph{MOST}}
\newcommand{\corot} {\emph{CoRoT}}

\newcommand{\kepler} {\emph{Kepler}}
\newcommand{\tess} {\emph{TESS}}

\newcommand{\cestam} {{\sc cestam}}
\newcommand{\filou} {{\sc filou}}

\title[Low-order \Dnu-\rhom\ for moderately-rotating \ds]{Study of the low-order \Dnu-\rhom\ relation for moderately-rotating \dss\ and its impact on their characterisation}

\author[J.E. Rodr\'{\i}guez-Mart\'{\i}n et al.]{
J.~E. Rodr\'{\i}guez-Mart\'{\i}n,$^{1,2}$\thanks{E-mail: julioeroma@correo.ugr.es, julioeroma@iaa.es}
A. Garc\'{\i}a Hern\'andez$^{1,2}$\thanks{E-mail: agh@ugr.es}
J.~C. Su\'arez$^{1,2}$\thanks{E-mail: jcsuarez@ugr.es}
and J. R. Rod\'on$^2$
\\
$^{1}$Departamento de F\'{\i}sica Te\'orica y del Cosmos, Universdidad de Granada, Campus de Fuentenueva s/n, 18071, Granada, Spain\\
$^{2}$Instituto de Astrof\'{\i}sica de Andaluc\'{\i}a (CSIC). Glorieta de la Astronom\'{\i}a s/n. 18008, Granada, Spain
}

\date{Accepted XXX. Received YYY; in original form ZZZ}

\pubyear{2019}

\begin{document}
\label{firstpage}
\pagerange{\pageref{firstpage}--\pageref{lastpage}}
\maketitle

\begin{abstract}

The large separation in the low radial order regime is considered as a highly valuable observable to derive mean densities of \dss, due to its independence  with rotation. Up to now, theoretical studies of this \Dnu-\rhom\ relation have been  limited to 1D non-rotating models, and 2D pseudo-evolutionary models. The present work aims at completing this scenario by investigating quantitatively the impact of rotation in this relation on a large grid of 1D asteroseismic models representative of \dss. These include rotation effects on both the stellar evolution and the interaction with pulsation. This allowed us to compute the stellar deformation, get the polar and equatorial radii, and correct the stellar mean densities. We found that the new \Dnu-\rhom\ relation for rotating models is compatible with previous works. We explained the dispersion of the points around the linear fits as caused mainly by the distribution of the stellar mass, and partially by the evolutionary stage. The new fit is found to be close to the previous theoretical studies for lower masses ($1.3-1.81\,\mathrm{M}_{\odot}$). However, the opposite holds for the observations: for the higher masses ($1.81-3\,\mathrm{M}_{\odot}$) the fit is more compatible with the empirical relation. To avoid such discrepancies, we provided new limits to the fit that encompass any possible dependency on mass. We applied these results to characterise the two well-known \dss\ observed by CoRoT, HD\,174936 and HD\,174966, and compared the physical parameters with those of previous works. Inclusion of rotation in the modelling causes a tendency towards greater masses, radii, luminosities and lower density values. Comparison between \Dnu\ and Gaia's luminosities also allowed us to constraint the inclination angles and rotational velocities of both stars. The present results pave the way to systematically constrain the angle of inclination (and thereby the actual surface rotation velocity) of \dss.

\end{abstract}




\section{Introduction}

$\delta$ Scuti stars  are the perfect laboratory to study the effects of rotation. They are stars of late A or early F spectral type, so they are usually rapid rotators \citep{Royer2007}. Their mass is 1.5-2.5 $\text{M}_{\odot}$ and they belong to the Population I stars (except for the subgroup SX Phe, which belong to the Population II group). They are placed where the main sequence crosses the classic instability strip. The oscillation mechanisms are mainly maintained by  the $\kappa$ mechanism (see for example \citealt{Asteroseismology2010}). They show radial and non radial-modes. They are generally low order p modes with periods from 18 min up to 8 hr. Observed amplitudes go from mmag up to the tenths of magnitude \citep[see, for example,][for a review on their characteristics]{Rodriguez&Breguer2001}.\par

In the last few years, thanks to the technological advances (mainly the space satellites) deeper studies have been done. \cite{Uytterhoeven2011} studied a sample of more than 700 stars with spectral type A-F, observed by the Kepler satellite, finding that about 23\% of them showed a hybrid $\gamma$ Dor - $\delta$ Scuti behaviour, since they showed p and g modes simultaneously. \cite{Murphy2019} used the parallaxes measured by GAIA to obtain precise values of the luminosity of 15000 stars observed with Kepler, finding that not all stars in the classic instability strip are pulsators, only about a 60\%. On the other hand, it has been found that the $\kappa$ mechanism is not the only responsible of maintaining the oscillations of the $\delta$ Scuti star. Coupling between oscillation and convection has also an important role, as shown by \citet{Dupret2005}, proving that more modes could be excited and suggesting that the $\delta$ Scuti and the $\gamma$ Dor belong to a same variable star type. Moreover, \cite{Xiong2016} showed that the turbulent motions of convection have to be also taken into account to explain the observed frequency ranges. All this only complicates the interpretation of the oscillation spectra.\par

In the era of space missions like \most\ \citep{most}, \corot\ \citep{Baglin2006}, \kepler\ \citep{Gilliland2010}, and now \tess\ \citep{Ricker2009b} thanks to ultra-precise photometric time series, periodic patterns could be detected in the p-mode frequency spectra of \dss\ \citep[see e.g.][from now on GH09 and GH13, respectively]{GH2009, GH2013}. These patterns were also predicted theoretically \citep{Reese2017, Ouazzani2015} and were found to be compatible with a large separation \citep[\Dnu][ from now on S14]{Suarez2014} since it is related to the stellar mean density. This relation was also confirmed for higher rotation rates using 2D models \citep{Reese2008,Mirouh19}. It  was empirically proved using binary systems with a \ds\ component \citep[][ from now on GH15]{GH2015}. Later on, \citet[][ from now on GH17]{GH2017} showed that it is possible to accurately determine surface gravity of those stars from the the \Dnu-\rhom\ relation and a measurement of the parallax. Thanks to all this progress, it has been possible to perform the first mutil-variable analysis on observed seismic data \citep{Moya2017}, which is an important step toward massive seismic studies of A-F stars.\par


Despite this progress, the $\Delta\nu-\bar{\rho}$ relation is still poorly studied.The previous works mentioned above either used non-rotating (S14) or \textbf{static} 2D equilibrium models \citep{Reese2008, Mirouh19} to derive the relation. \textbf{Although stellar evolution is mimicked in the latter by considering models with different core hydrogen abundances, this remains a crude approximation compared to what is achieved in 1D stellar evolution codes. Accordingly, the conclusions of these works remain limited.}  In this context, here we explore the large separation-mean density relation using \textbf{an extensive} grid of rotating models and analyse its behaviour with the physical stellar parameters.

The paper is organised as follows: in Section~\ref{sec: methodology} we explain the methodology of the work, including the characteristics of our asteroseismic models. In Section~\ref{sec:relation} we revisit the \Dnu\ vs \rhom\  relation to study how it is modified by rotation, to what extent, and the implication on asteroseismic determination of stellar magnitudes. In Section~\ref{sec:characterisation} we applied the methodology to two \dss\ that were previously studied without including rotation effects. Finally, conclusions are outlined in Section~\ref{sec:conclusions}.\par


%
\begin{table}
    \centering
    \caption{Input parameters for \cestam. $M$ is the stellar mass (in solar masses), $\alpha$ is the convective efficiency of the mixing length theory, [Fe/H] is the metallicty (in dex), ov is the overshooting parameter, $p$ the initial rotation period (in days) and $\tau$ the time the protoplanetary disk corrotates with the star (in days). }
    \begin{tabular}{c|c|c}
        \hline
        Input parameter &  Range & Step \\  
        \hline 
         $M$ & [1.30, 3.00] M$_{\odot}$ & 0.05 M$_{\odot}$ \\
         $\alpha$ & 1.64 & 0 fixed \\
         \ [Fe\/H] & [-0.4, 0.2] & 0.1 \\
         ov & 0.0 & fixed \\
         $p$ & [5, 7] & 1 \\
         $\tau$ & 5 & fixed
    \end{tabular}
    \label{tab:CESTAMinput}
\end{table}
%

%
%

\begin{figure*}
    \centering
    \includegraphics[width=\columnwidth]{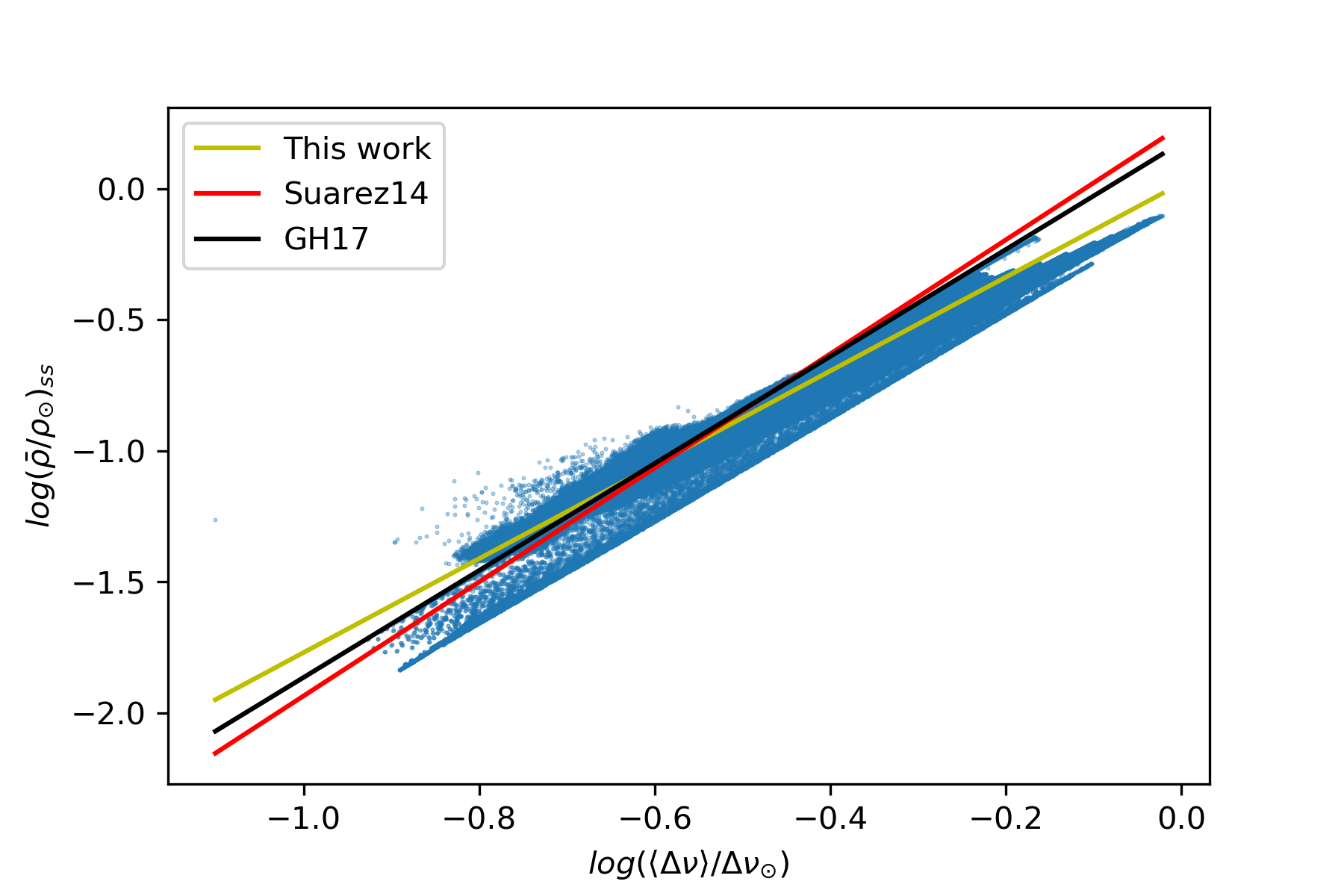}
    \includegraphics[width=\columnwidth]{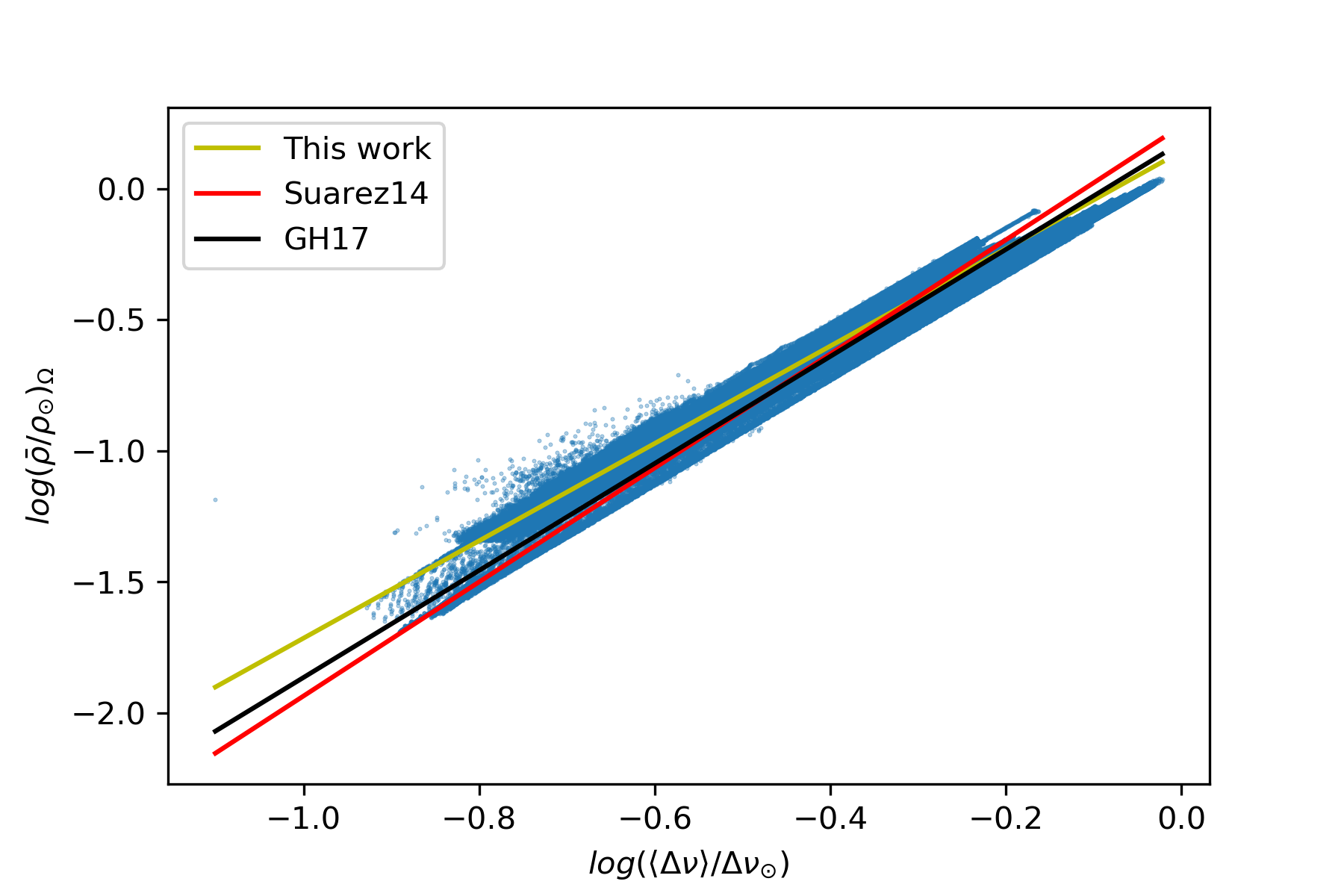}
    \caption{Mean density as a function of the average large separation, both in solar units and in logarithmic scale. The fit obtained in this work is represented, as also the linear fits obtained in previous works for comparison. Left panel shows the relation for the spherical symmetric model and right panel for the spheroid (see text for details).}
    \label{fig:CEFIRO_fit}
\end{figure*}

\section{Methodology} \label{sec: methodology}

The main objective of this work is to theoretically study the impact of rotation on the relation \Dnu-\rhom\ for A-F stars. To do so, we first need to build a grid of stellar asteroseismic models (stellar structure and oscillations) representative of those stars, similar to the one used by S14 but including rotation during the evolution. With those models, we followed S14's methodology to compute the large separations, which are then paired with the stellar mean density obtained directly from the structure models. We then compare the new $\bar{\rho}-\Delta \nu$ relation obtained in this work with those obtained previously without including rotation effects \citep{Suarez2014, GH2015, GH2017}.

\subsection{The equilibrium models}\label{ssec: eqmodels}

We built a grid composed of more than half a million stellar equilibrium models computed with the \cestam\ code \citep{Marques2013} distributed in 663 evolutionary tracks representative of intermediate-mass stars following in a similar manner as in S14. Models are evolved from the early pre-main sequence (assuming a proto-star with a proto-planetary disk) up to the sub-giant branch. We constructed the grid by varying three input parameters: mass, metallicity, and initial rotation velocity. This latter is obtained varying the initial rotation period of the protostar and the time ($\tau$) during which the star and the protoplanetary disk are locked. In contrast to S14, we fixed the mixing length convection efficiency to $\alpha = 1.65$ with no overshoot since we consider rotationally induced mixing as prescribed by \citet{Zahn1992} and a further refinement by \citet{Mathis2004} for the radiative zones of the stellar interior \citep[more details in][]{Marques2013}.\par

In order to ensure a correct computation of the adiabatic oscillations, we imposed a number of shells in which the stellar structure is radially distributed to be above 2000 approximately, as suggested by the ESTA/CoRoT working group \citep{Moya2008, Lebreton2008}. Moreover, the stellar models were computed to rotate with $\Omega/\Omega_\text{C} \lesssim 0.7 $, the vast majority below 0.5.\par

\subsection{The stellar oscillations}\label{ssec:oscillations}

Stellar adiabatic oscillations were computed using the code \filou\ \citep{Suarez&Goupil2008}, which computes adiabatic oscillations using a perturbative approximation. It takes into account the stellar distortion due to the centrifugal force in the oscillation frequency computation. Moreover, the code corrects the oscillation frequency for the effects of rotation up to the second order, including near-degeneracy effects \citep[see][for more details]{Suarez2006}.\par

Theoretical oscillation modes were computed for each of the equilibrium models of the grid with spherical degree in the range $0 \le \ell \le 2$, from a few low-order g modes up to the cut-off frequency, which, for these stars is found in the pressure (p) modes domain.

\section{The $\Delta \nu-\bar{\rho}$ relation} \label{sec:relation}

\subsection{The large separation} \label{ssec:large separation}

The relation between the large separation and the mean density of \dss\ predicted in S14 held even for rotating models because a significant effect on $\Delta\nu/\sqrt{{\bar\rho}}$ with the distortion of the star is not expected  up to 80\% of the break up rotation frequency \citep{Reese2008}. Therefore, a similar relation should be found when rotating models are considered. Following S14, we compute for each model the large separation for each oscillation mode as: 
\begin{equation}
   \Delta \nu_{\ell}= \nu_{n+1,\ell}-\nu_{n,\ell},\label{eq.dnu} 
\end{equation}
where $\nu_{n, l}$ is the frequency of the mode of radial order $n$ and spherical degree $\ell=0,1,2\, ($m=0$)$. To enhance the pattern related with the large separation, we restricted the calculation of \Dnu\ to the range in which \dss\ pulsate, i.e. $2 \le n \le 8$. The actual  $\Delta\nu_{\ell}$ for each $\ell$ is calculated as the median of the individual $\Delta\nu_{n,\ell}$ values. We choose the median instead of the average because it is more stable in case an avoided crossing blurs \Dnu. In practice, we do not know (by now) which modes are contributing to the observed \Dnu. We computed the median (instead of the average, as it was adopted in S14) of all $\Delta\nu_{\ell}$ as a better approximation to the observed large separation. No significant differences between $\Delta\nu_{\ell}$ and the median \Dnu\ is found. Only a slightly larger dispersion was found for $\ell=2$, which is explained by the avoided-crossing phenomenon, not fully avoided by considering the median. In any case, such a dispersion, as happened in S14, was found to be statistically negligible. Therefore, in what remains of paper we will use only the average of the large separations $\left \langle \Delta \nu \right \rangle $.\par
\begin{figure*}
    \centering
    \includegraphics[width=\columnwidth]{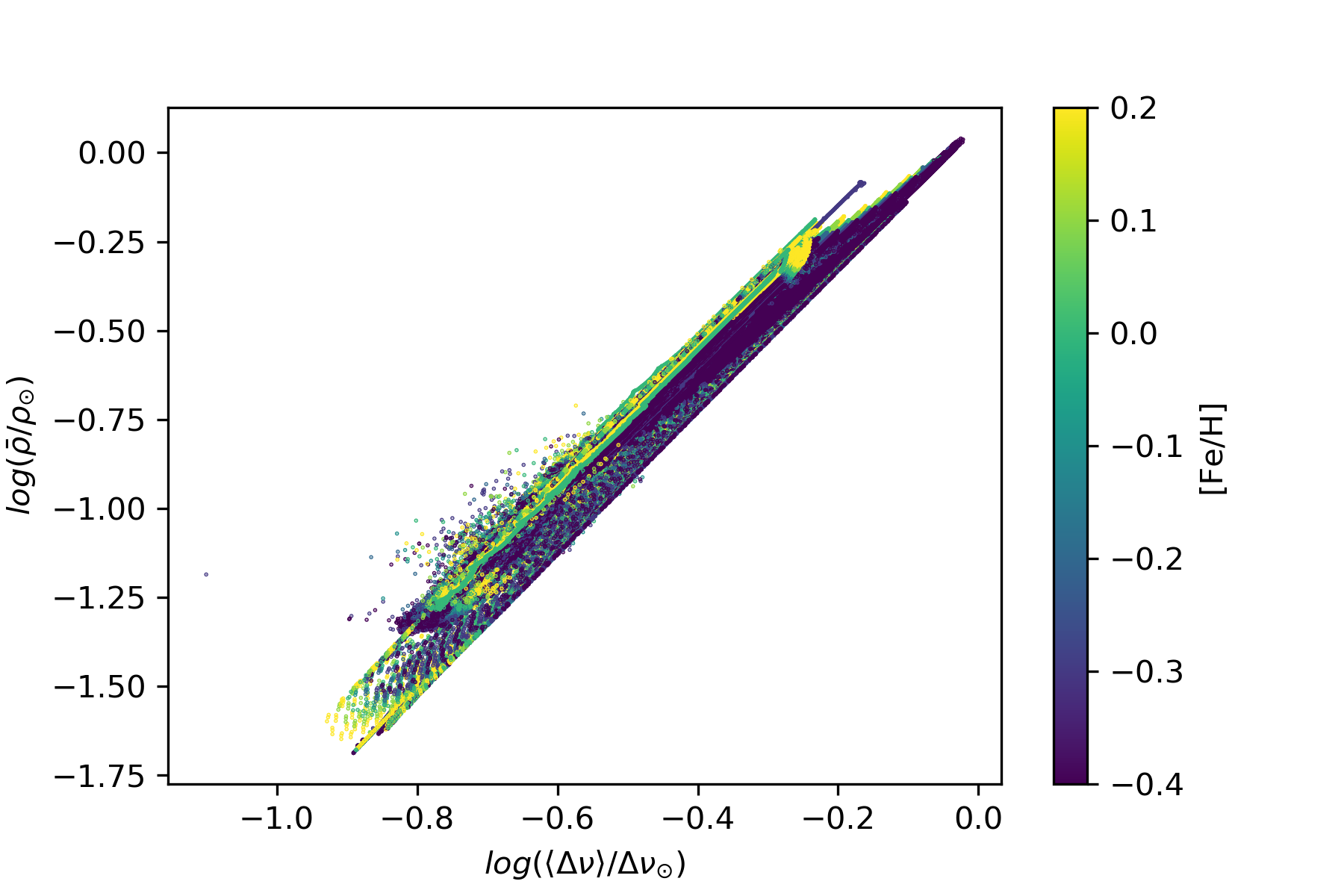}
    \includegraphics[width=\columnwidth]{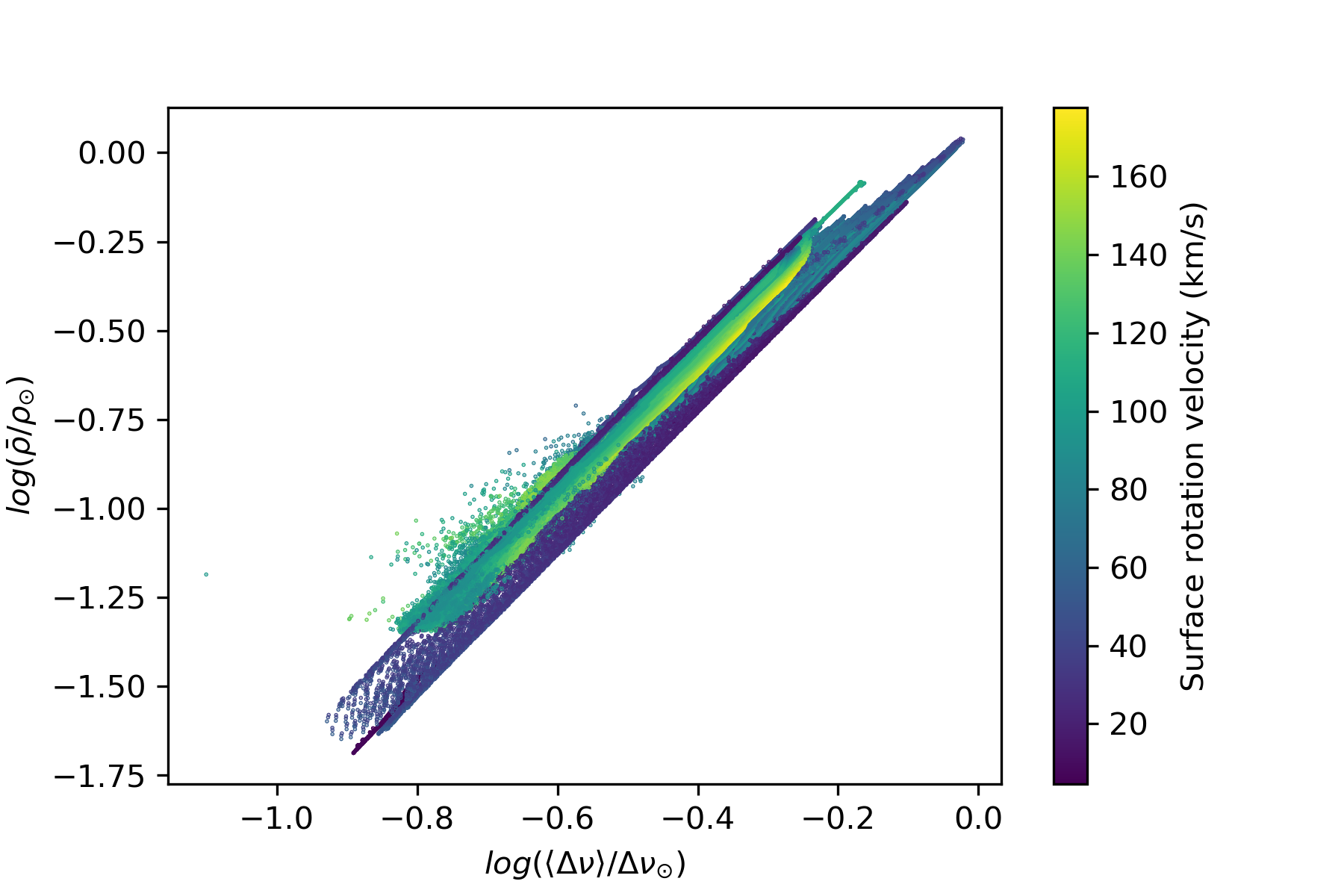}
    \includegraphics[width=\columnwidth]{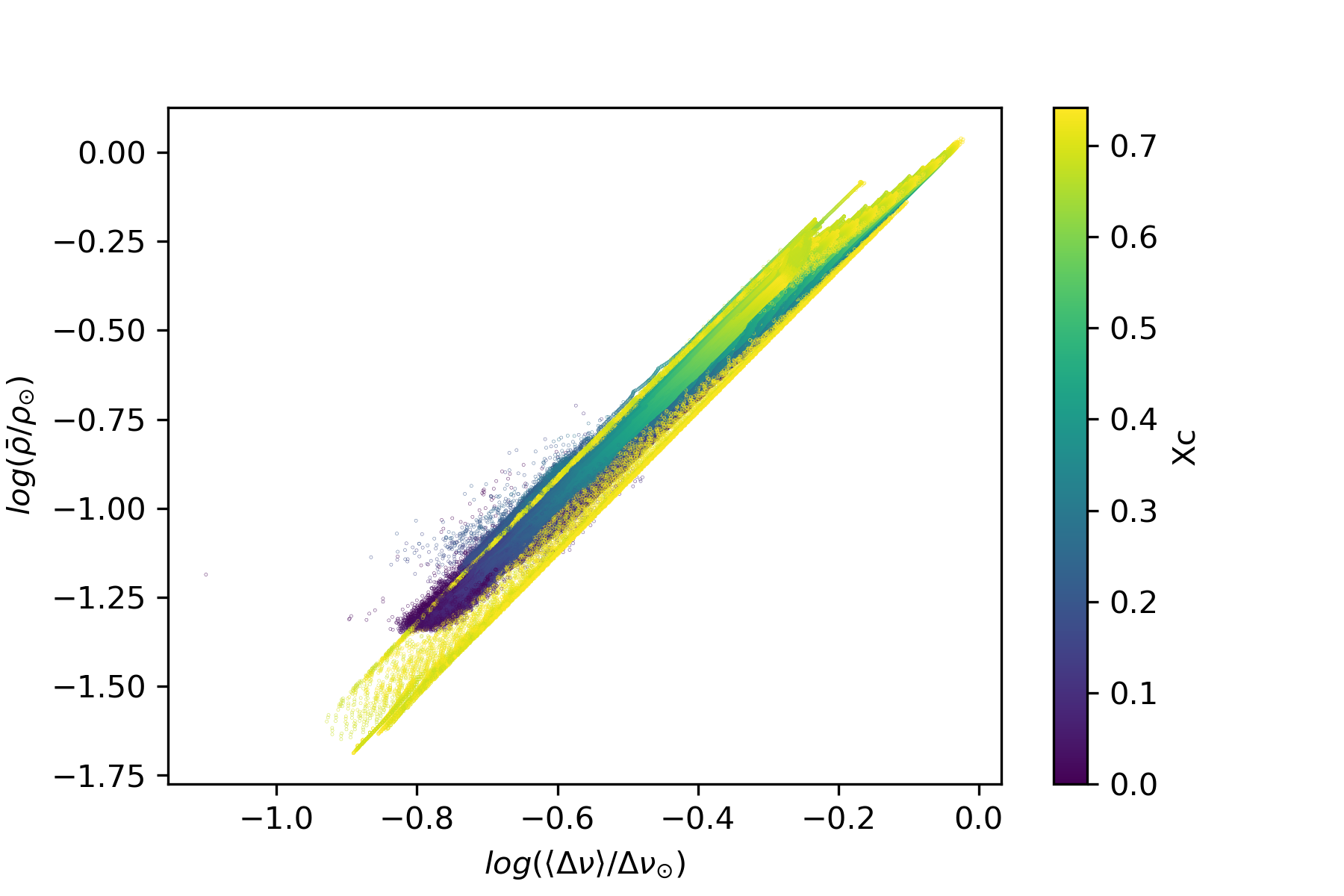}
    \includegraphics[width=\columnwidth]{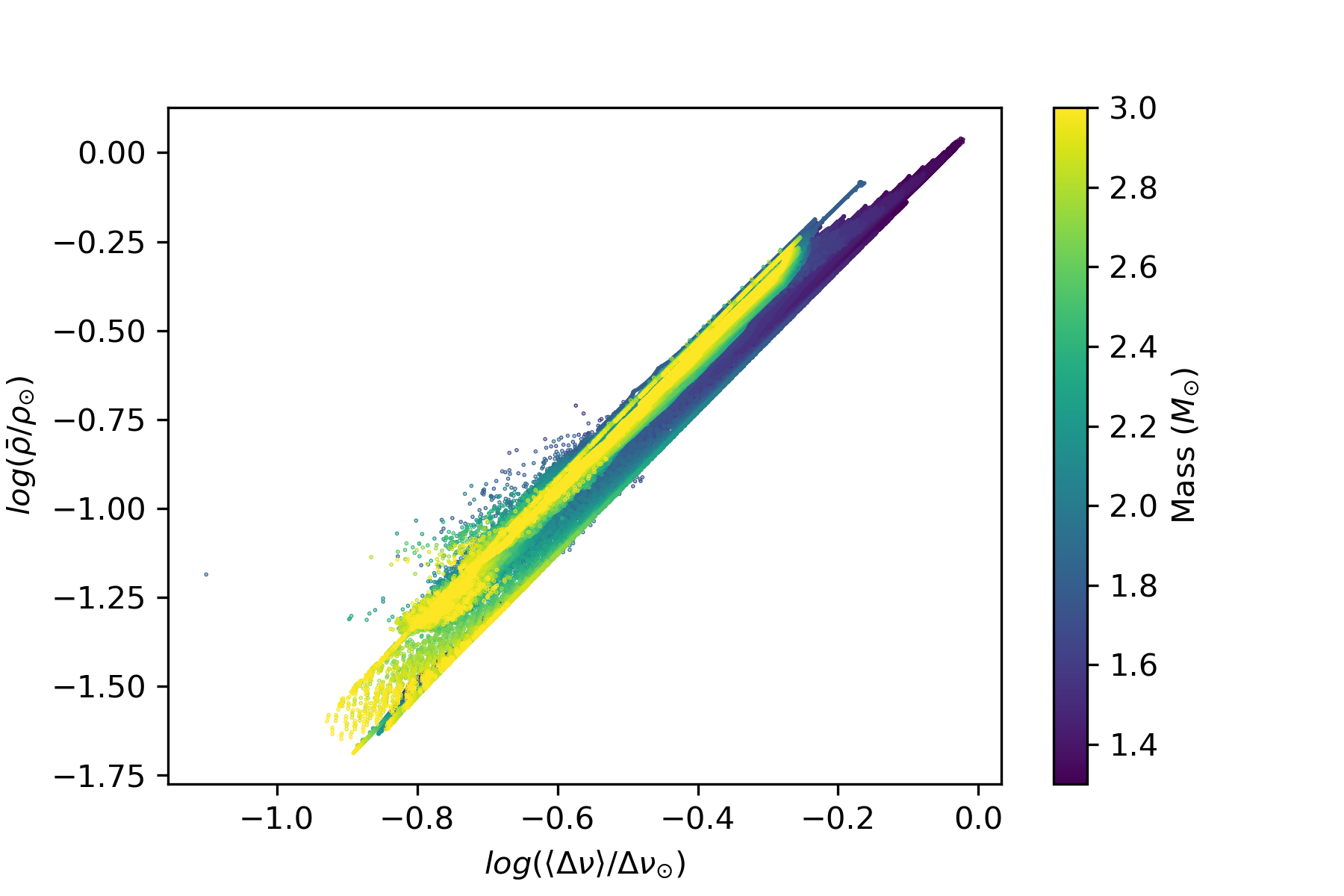}
    \caption{Mean density as a function of the large separation, both in logarithmic scale and normalised to the solar values. The different panels show the variation (as a colour scale) of the metallicity (top, left), the central hydrogen fraction (bottom, left), the surface rotation velocity at the equator (top, right), and mass (bottom, right).}
    \label{fig:dnurhom.param}
\end{figure*}
\subsection{Mean density corrected for the effect of rotation} \label{ssec:mean density}

In order to properly assess the impact of rotation on the \Dnu-\rhom\ relation, it is important to consider the stellar deformation since it modifies the volume of the star. It is well known, that the shape of a rotating star under the assumption of shellular rotation can be well approximated by a Roche potential model (equipotential surfaces). Nevertheless, a spheroid is a reasonable approximation up to almost 75\% of the breakup rotational velocity \citep[see, e.g., Fig.~32 in ][]{Paxton2019}. Since all our models are well below this value, we used the volume of a spheroid to compute the mean density. To that end, we needed to obtain the polar and equatorial radii from the parameters provided by {\sc cestam}. So first, we need to understand what are those parameters.\par

Stellar structure quantities in {\sc cestam} are defined as a mean variable over an isobar plus a perturbation term of the form \citep{Marques2013}:
\begin{equation}
   f(p,\theta) = \Bar{f}(p)+\Tilde{f_2}(p)\,P_2(\cos{\theta}), \label{eq:cestam variables} 
\end{equation}

where $P_2(\cos{\theta})$ is the second-order Legendre polynomial, $p$ is the pressure, and $\theta$ is the colatitude. Moreover, $r$ is defined as the mean radius of the isobar. This means that {\sc cestam}'s stellar radius indeed corresponds to the radius of a spherical symmetric model (hereafter, $R_\text{SS}$) when $P_2(\cos{\theta})=0$, that is, $\sin{\theta}^2=2/3$. Thus, following \citet{PerezHernandez1999} and assuming that the polar radius, $R_\text{p}$, barely change with rotation for the same stellar mass, we can derive it as:
\begin{equation}
   R_\text{p}=\frac{R_\text{SS}}{1+\frac{\Omega^2 R^3_\text{SS}}{3GM}},\label{eq:polar radius} 
\end{equation}

where $G$ is the gravitational constant, $M$ is the stellar mass and $\Omega$ is the angular rotational velocity.\par

To derive the equatorial radius, $R_\text{e}$, we need to account for the stellar deformation. This is done using the fraction of critical rotation, $\omega=\Omega/\Omega_\text{C}$.  Here, $\Omega_\text{C}$ is the critical angular rotation velocity, i.e. the velocity when the centrifugal and gravitational forces are balanced. At this point, $R_\text{p}=2/3R_\text{e}$ \citep[see, e.g.,][]{Paxton2019}. We can thus relate the critical rotational velocity with the polar radius as follows:

\begin{equation}
   \Omega^2_\text{C}=\frac{8GM}{27R^3_\text{p}},\label{eq:Omega_C} 
\end{equation}
which allows us to relate both the equatorial and polar radii \citep[see][for details]{Paxton2019} as:
\begin{equation}
   \frac{R_\text{e}}{R_\text{p}} = 1+\frac{\omega^2}{2}.\label{eq:Re/Rp} 
\end{equation}
We can now calculate the \emph{correct} mean density of the models (as ellipsoids), from these radii as:
\begin{equation}
   V = \frac{3}{4\pi} \frac{M}{R_\text{e}^2 R_\text{p}}.\label{eq:newrho}
\end{equation}

\subsection{Discussion}\label{ssec:discussion}

Using {\sc cestam}'s output and computing the volume of a spherically symmetric model directly with the radius it provides can lead to noticeable differences (see Fig.~\ref{fig:CEFIRO_fit}). \emph{Correct} mean densities are systematically lower (larger volumes), and it is noticeable that the \Dnu -\rhom\ relation is tighter, i.e. the dispersion of points is reduced as compared with the spherically symmetric density. \par

Such effects modify slightly the fit of the overall behaviour of points in the \Dnu - \rhom\ diagram:
\begin{figure*}
    \centering
    \caption{Mean density as a function of the large separation, both in logarithmic scale and normalised to the solar value, for the 10 mass intervals considered in this work. The central hydrogen mass fraction is represented in colour scale. For comparison, S14's and GH17's fits are also displayed.}
    \includegraphics[width=1.4\columnwidth]{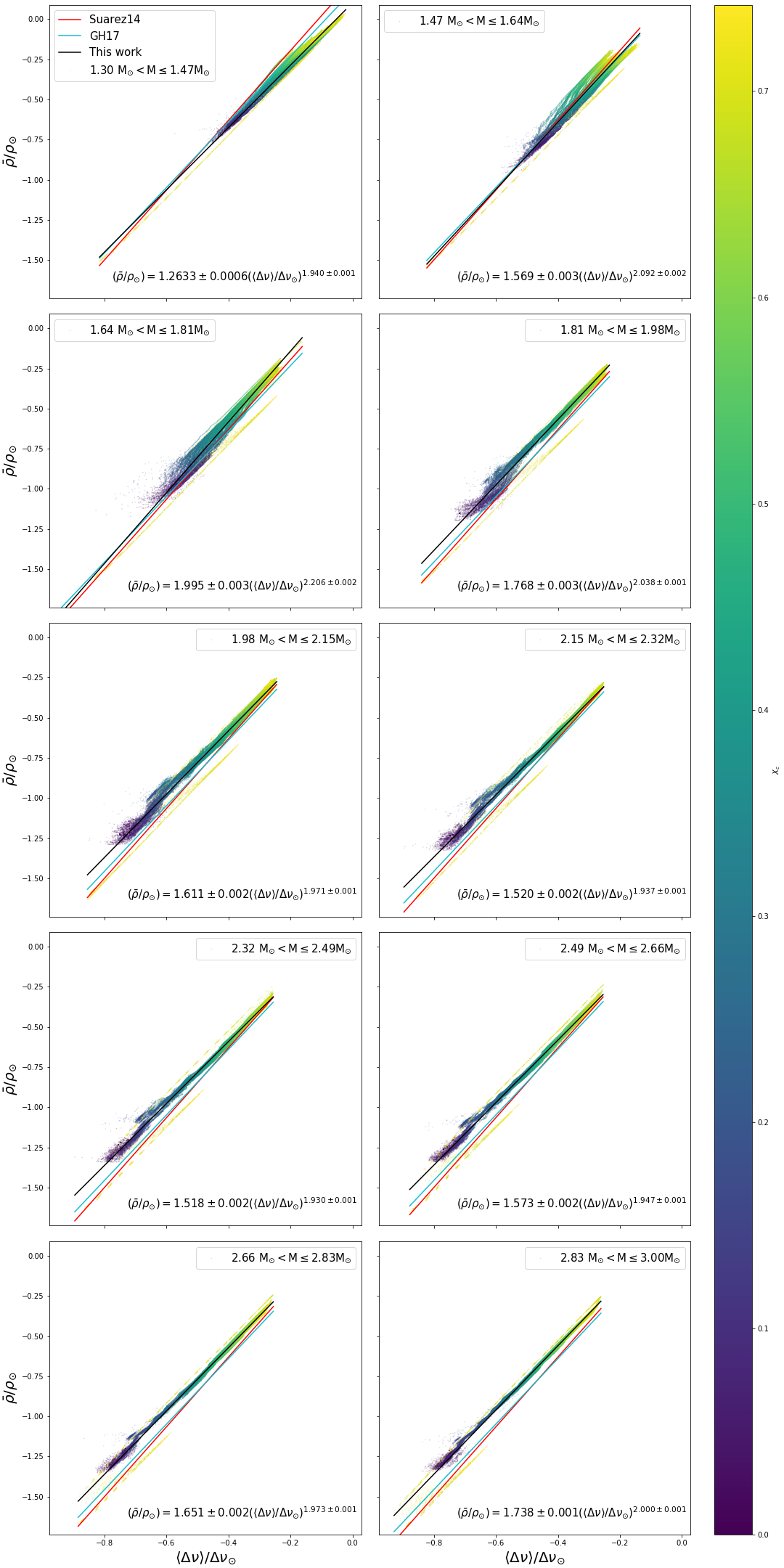}
    \label{fig:fitintervals_complete}
\end{figure*}
\begin{align}
\left(\frac{\bar{\rho}}{ \rho_{\odot}}\right)_{\Omega} &= 1.3867^{+0.0006}_{-0.0006}\left( \frac{< \Delta \nu  > }{ \Delta \nu_{\odot}} \right)^{1.8561^{+0.0004}_{-0.0004}} \\
\left(\frac{\bar{\rho}}{ \rho_{\odot}}\right)_{\mathrm{ss}} &= 1.0481^{+0.0005}_{-0.0005}\left( \frac{< \Delta \nu  > }{ \Delta \nu_{\odot}}\right)^{1.7894^{+0.0005}_{-0.0005}}
\label{eq:fitcomp}    
\end{align}
where subscript SS stands for the spherical symmetric case and $\Omega$ for the spheroid. Both fits have a Pearson correlation parameter above 0.97. The fact that fits are closer to empirical predictions (GH17) for deformed models give us more confidence in our account for rotation in overall asteroseismic modelling. In addition, it illustrates the importance of properly considering the stellar shape when computing structure quantities of a deformed star. Hereafter, deformed models (non-spherical density models) are considered.\par

As in S14, we obtain a clear linear relation (in logarithmic scale) between the large separation and the mean density, although with significantly different fitting parameters. We found differences of about 20\% for the exponent and 70\% for the multiplicative factor with respect to S14's fit and around 10\% for the exponent and 50\% for the multiplicative factor compared to GH17's. In order to understand such differences, we studied the main \emph{suspects}: metallicity, mass and evolutionary stage distribution. In the present work, an additional parameter comes into play: the stellar surface rotation. \par

The distribution of those parameters in the \Dnu -\rhom\ diagram shows, as expected, a behaviour similar to what was found in S14 (Fig.~\ref{fig:dnurhom.param}). Metallicity is uniformly distributed with no specific trend. Rotational velocity shows a distribution that follows the evolution of the star with higher velocity for younger models (as expected) mainly located in the bunch of points around the middle of the panel (green/yellow dots). Regarding the central hydrogen fraction ($X_{\mathrm{c}}$), we found bands placed along the fit direction that reflects the evolutionary stage of the models, yet not explaining the major dispersion of points in the perpendicular direction. Finally, the only parameter of the four here analysed that does not depend on the stellar evolution is the mass, which shows gradient of bands perpendicular to the fit's direction. This is thus the main contributor to the width of the \Dnu -\rhom\ diagram when all the models of the grid are displayed.\par

Because of their importance, we decided to go deeper into the analysis of both the $X_{\mathrm{c}}$ and $M$ parameters combined. The bottom objective is to better understand the dependence of the fit with mass and Xc. For this purpose, we divided the model grid into ten equal mass range intervals on which we performed a linear regression to the \Dnu-\rhom\ relation in logarithmic scale (see Fig.~ \ref{fig:fitintervals_complete}).\par

\begin{table*}
    \centering
    \caption{Observables used for the characterisation of HD\,174636 and HD\,174966. The second row represents the photometric observables of HD\,174966, meanwhile the third represents the spectroscopic ones. References: $^1$\citet{Charpinet2006}, $^2$GH09, $^3$\citet{GAIAcol2018b}, $^4$\citet{Solano2005}, $^5$GH13.}
    \begin{tabular}{c|c|c|c|c|c|c}
        \hline
        Star & T$_\text{eff}$ (K) & log g & [Fe/H] & $\Delta \nu$ ($\mu$Hz) & L(L$_{\odot}$) & v$_\text{rots}$ (km/s) \\
        \hline    
        HD 174936 & $8000 \pm 200^1$ & $4.08 \pm 0.20^1$ & $-0.32 \pm 0.20^1$ & $52 \pm 10^2$ & [10.79, 11.15]$^3$ & $\ge 169.7^1$ \\
        HD174966 & 7637 $\pm$ 200$^{4}$ & $4.03 \pm 0.20^{4}$ & -0.11 $\pm$ 0.20$^{4}$ & 65 $\pm$ 1$^{5}$ & [10.73, 10.93]$^{3}$ & [135,178]$^{5}$ \\
        HD174966 & 7555 $\pm$ 50$^{5}$ & $4.21 \pm 0.05^{5}$ & -0.08 $\pm$ 0.10$^{5}$ & 65 $\pm$ 1$^{5}$ & [10.73, 10.93]$^{3}$ & [135,178]$^{5}$ \\
        \hline
    \end{tabular}
    \label{tab:observables}
\end{table*}

\begin{figure}
    \centering
    \includegraphics[width=\columnwidth]{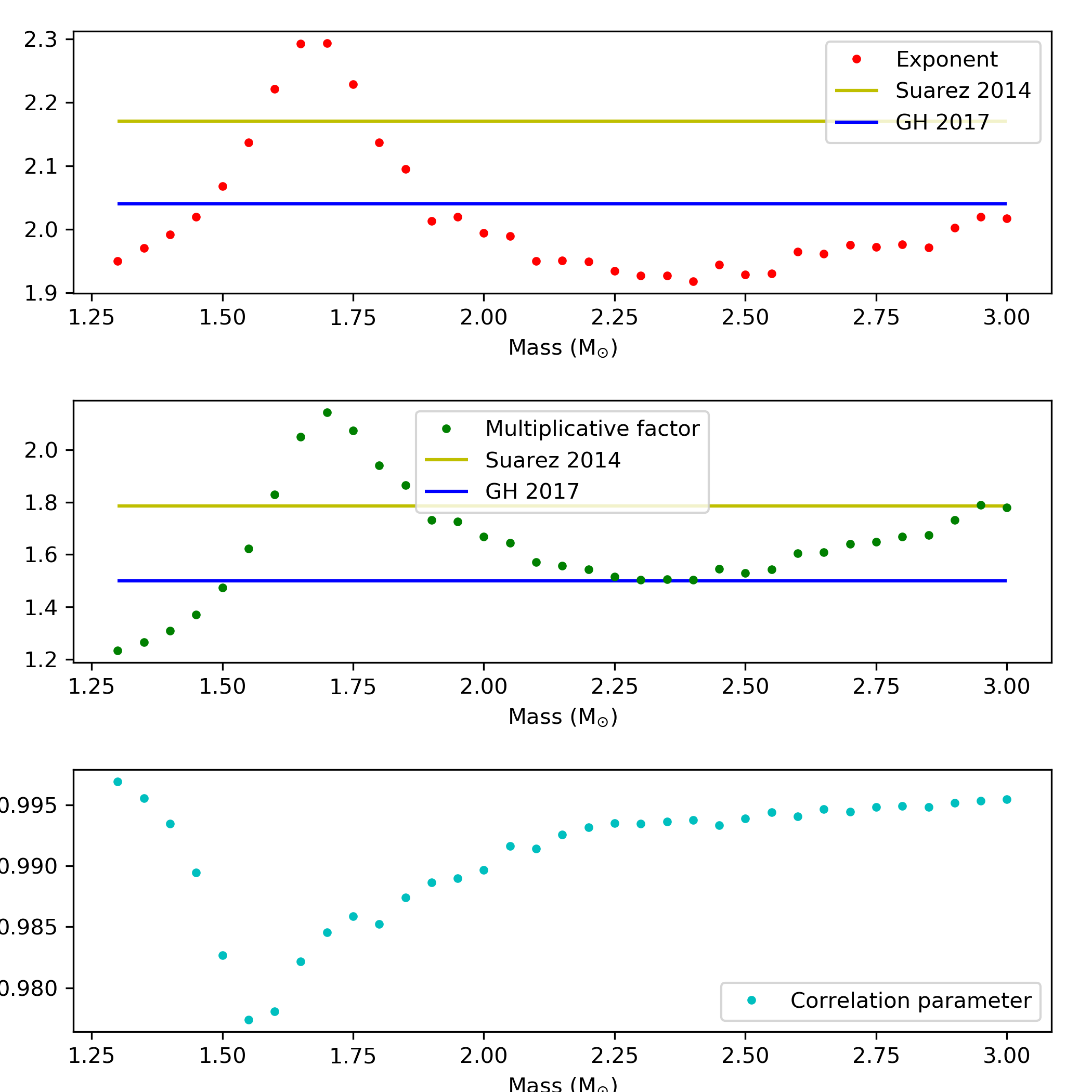}
    \caption{Upper panel: Value of the exponent of the fit as a function of mass. Results of previous works are represented for comparison. Middle panel: Value of the multiplicative factor of the fit as a function of mass. Results of previous works are represented for comparison. Lower panel: Correlation factor of each fit.}
    \label{fig:Fit_dependency}
\end{figure}

Notice that the highest mass range fit is the furthest away from the S14 and GH17 fits, whereas the lowest mass range are the closest to them. In between, there seems to be a \emph{cluster} of models (generally late main sequence and subgiant branch) that determine the main weight of the fit, which move from the lower-left to the upper-right parts of the panel.\par

In particular, for masses lower than 1.81 $\text{M}_{\odot}$, the differences in the exponent are of up to the 1\% with S14 and the 8\% with GH17. In the case of the multiplicative factor, these differences are up to the 55\% with S14 and the 33\% with GH17. However, when considering masses above 1.81 $\text{M}_{\odot}$ these differences become smaller with respect to GH17 but higher with respect to S14. Exponents were found to be lower by 12\% when comparing with those given by S14, and lower by 6\% compared with those by GH17. In the case of the multiplicative factor they are lower by 38\% and 18\%, respectively. The fact that the agreement in the multiplicative factor is always poorer was expected, since it is obtained via the independent term of the linear fit, and therefore it is a less precise value.\par

In addition, a subset of young models seems to populate the same region of the diagram whatever mass interval is considered, although with different sizes in the direction of the fit (yellow colour points in Fig.~\ref{fig:fitintervals_complete}). We identify these as models in the ZAMS, or near the PMS close to the ZAMS. This range of models are well described by S14 and GH17 fits, and are compatible with the relations found with 2D pseudo-evolutionary models \citep{Mirouh19}.\par

We then reduced the size of the mass \emph{buckets} to the individual masses of the models in order determine the fit's actual dependence with mass. We applied thus the linear regression fit to all the models of each mass (Fig.~\ref{fig:Fit_dependency}). We found similar behaviour in both fitting parameters: there is an initial increase, reaching a peak at the same mass (about 1.70 M$_{\odot}$), and then decreasing until a plateau. In the case of the exponent this plateau appears at a value close to that of the lowest mass (1.25~M$_\odot$) and corresponding to the minimum value of the exponent. Meanwhile, the multiplicative factor reaches a value that corresponds to the GH17's relation. The lowest value of the correlation parameter is around 0.970, meaning that all fits are very solid. The difference between the maximum and minimum value are about a 20\% for the exponent and 75\% for the multiplicative factor.\par

\begin{figure}
    \centering
    \includegraphics[width=\columnwidth]{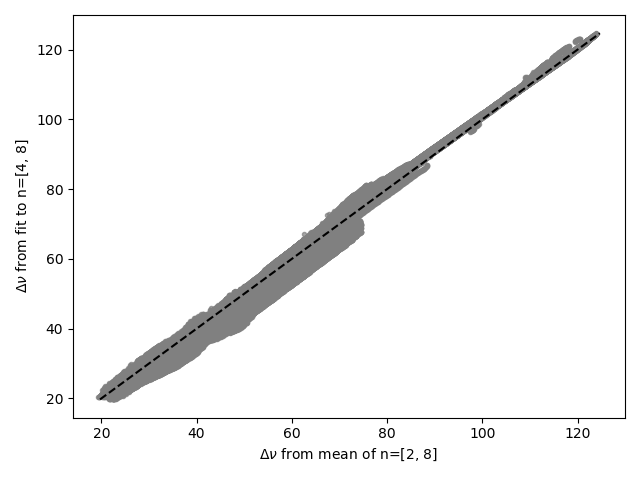}
    \caption{Comparison between $\Delta\nu$ obtained from our mean value and those obtained from fitting to $\nu = \Delta\nu(n+\varepsilon)$. See details in text.}
    \label{fig:Fit_comparison}
\end{figure}

\begin{figure*}
    \centering
    \includegraphics[width=\columnwidth]{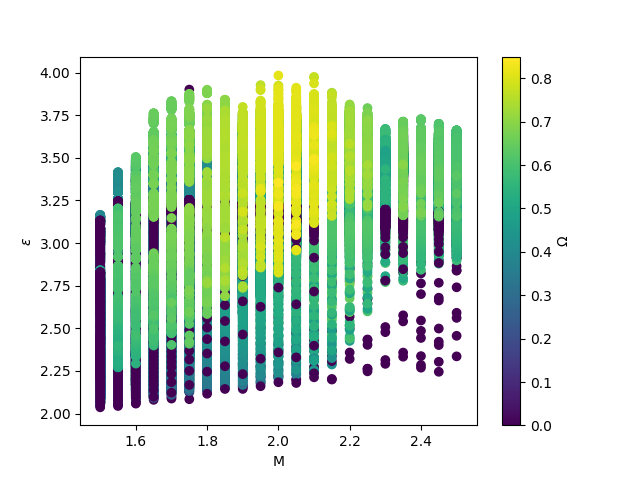}
    \includegraphics[width=\columnwidth]{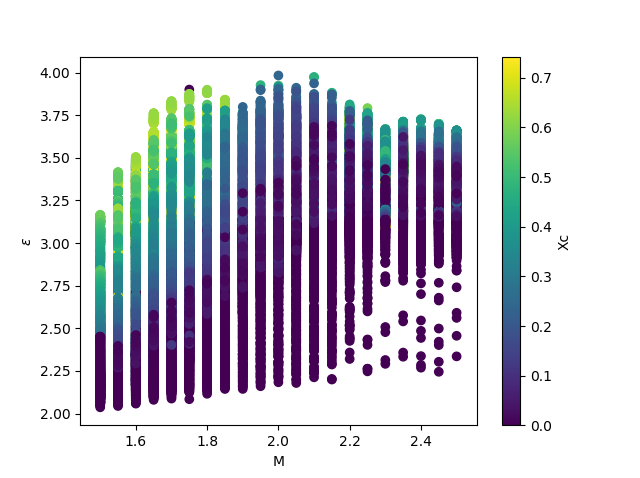}
    \caption{Left: $\varepsilon$ as a function of mass. Colour code indicates the value of the rotation rate. Right: same as left, but colour code indicates the fraction of central Hydrogen.}
    \label{fig:Epsilon_parameters}
\end{figure*}

It is worth noticing that the range of masses studied in S14 was 1.25-2.20 M$_{\odot}$. This means that the peak found in Fig.~\ref{fig:Fit_dependency} takes a greater weight when fitting the models, and this would explain the greater values found in both fitting parameters of S14. On the other hand, in GH17 only three out of eleven stars had a mass below 1.81 M$_{\odot}$, meaning that most of the fitted stars fall in the plateau zone, which explains the significant resemblance with its value.\par

In order to get rid of the above dependency on the mass, and minimise the impact of the different number of models in different evolutionary stages we calculated the mean values of the exponent and the multiplicative factor for all the masses:
\begin{equation}
\bar{\rho} / \rho_{\odot} = 1.6^{+0.5}_{-0.4}(\left < \Delta \nu \right > / \Delta \nu_{\odot})^{2.02^{+0.10}_{-0.10}},
\label{eq:fit}
\end{equation} 
Then, the uncertainty in the exponent is simply the standard deviation and the multiplicative factor will go from the minimum to the maximum value, assuring the full coverage of the grid.

\subsection{Application of a least-square fit to calculate $\Delta\nu$}

We investigated the impact of using the average of $\Delta\nu_{\ell}$ on the fit and thereby potentially on the conclusions. To do so, we compared our fits with those obtained using the generalised asymptotic formula (for $n>>\ell$): $\nu = \Delta\nu(n+\varepsilon)$. Here, we cannot assume the asymptotic regime, so we use the phase $\varepsilon$ which accounts for the deviation of that regime for frequencies of a given mode degree $\ell$ \citep[see details in][]{Bedding2020}. To properly compare with that paper the fit is calculated considering only $\ell =0$ frequencies with radial orders in the range $n=[4,8]$. As can be seen in Fig.~\ref{fig:Fit_comparison} both the average and linear square fit are compatible, with no significant difference that may modify the conclusions of the present work. This result was \textbf{somewhat} expected since both $n$ ranges overlap significantly.\par

Moreover, we also investigated the dependency of $\varepsilon$ with other parameters that might influence it, like the impact of evolution \citep[as suggested by][]{Bedding2020}. Interestingly we found that the evolutionary state cannot be derived only from $\varepsilon$ as shown in right and left panels of Fig.~\ref{fig:Epsilon_parameters}. Notice the dependence with the mass and rotation rate in addition to stellar evolution, which implies that $\varepsilon$ is not an optimum parameter to derive stellar ages. The discrepancies with \citeauthor{Bedding2020}'s results might be due to the non-rotating models they used to calculate $\varepsilon$.\par

In any case, the results presented here are very preliminary and a more detailed study of the parameter is needed to derive any solid conclusion.

\section{Characterisation of two $\delta$ Sct stars: HD 174936 and HD 174966} \label{sec:characterisation}

\subsection{The data}\label{ssec:data}
HD 174936 and HD 174966 are $\delta$ Scuti stars \citep[see ][respectively]{Perryman1997, Lefevre2009} observed by the \textit{CoRoT} \citep{Baglin2006} and GAIA \citep{Perryman2003, GAIAcol2016, GAIAcol2018b} satellites. \textit{CoRoT} provided high precision photometry during 27.2 days and GAIA obtained very precise parallaxes from which accurate and precise luminosities were derived.\par

HD 174636, whose spectral type is A2, was asteroseismologically characterised by \citet{GH2009}. They obtained a total of 422 significant oscillation frequencies. That made this star one of the two $\delta$ Scuti stars with the largest number of detected frequencies at that time, two orders of magnitude greater than from the ground. Using a Fourier transform technique, they found a frequency spacing that was identified as a low order large separation.\par

Similarly, HD\,174966, with spectral type of A3, was studied for the preparation of the \textit{CoRoT} target sample \citep{Poretti2003}. Later on \citet{GH2013} used the data obtained from the 27.2 days of uninterrupted CoRoT observation (initial run) to extract 185 significant peaks. As in previous works, the authors assumed the frequency spacing as a large separation (later confirmed theoretically and empirically by S14, and S15 respectively) to discriminate between representative models of the star.\par

However, these stars were characterised using stellar grids that did not consider rotation. Our grid can give an idea of the impact of rotation when constraining stellar parameters asterosismologically. Moreover, luminosities from Gaia provide a new observable imposing additional constraints in the parameter space.\par

\begin{figure*}
    \centering
    \includegraphics[width=\columnwidth]{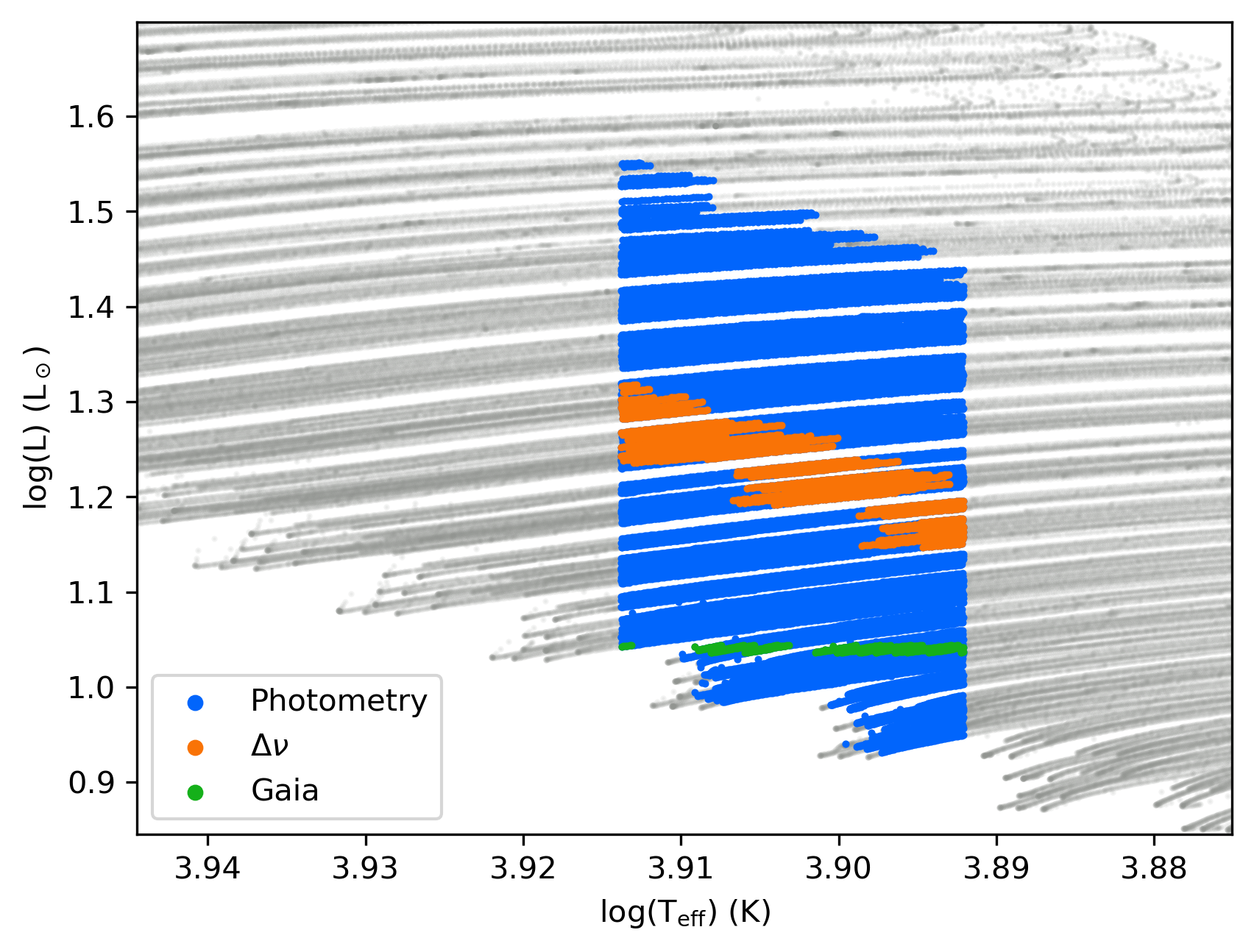}
    \includegraphics[width=\columnwidth]{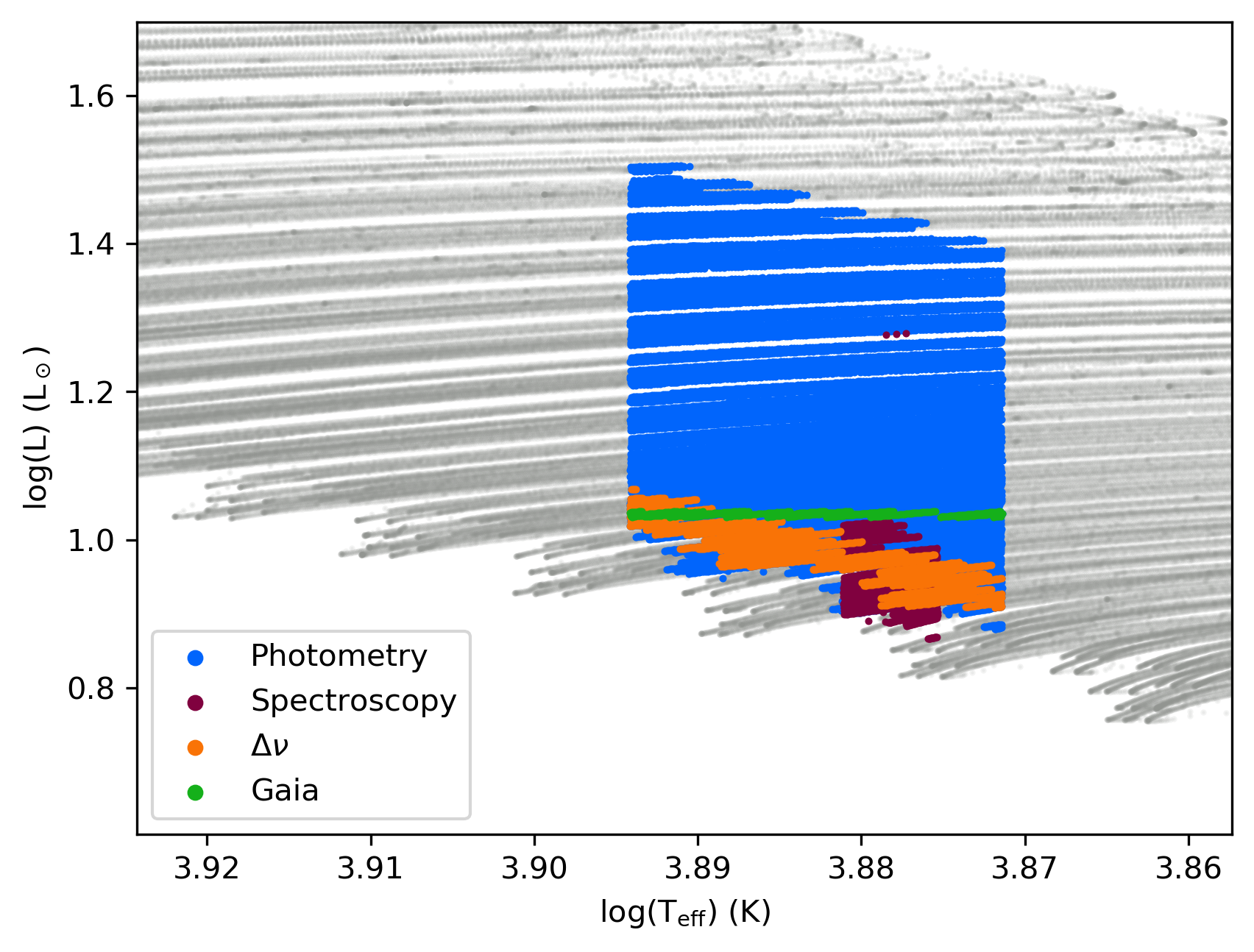}
    \caption{HR diagrams showing error boxes obtained for HD\,174936 (left) and HD\,174966 (right). Sources of uncertainty are depicted in different colours over the evolutionary tracks (grey).}
    \label{fig:stars}
\end{figure*}

\subsection{Characterisation}
The observables used for the characterisation (see Table~\ref{tab:observables}) were taken from GH09, GH13 and DR2 \citep{GAIAcol2018b}. HD\,174936 has only photometric parameters, whereas for HD\,174966 there exists both photometric and spectroscopic measurements. Generally, the star's angle of inclination is unknown or has very large uncertainty. We decided thus not to include the rotational velocity as a constraint for model discrimination.\par

The procedure followed, analogue to that followed in GH13, is also applied to the spectroscopic parameters of HD\,174966. First, we constrained models using the measured \teff, \logg\ and metallicities from photometry within the $1\sigma$ uncertainties. Then, we used \Dnu\ or Gaia's luminosity as discriminant. The discrimination showed a tendency towards greater values in mass and luminosity compared to the non-rotating case (see Tables~\ref{tab:HD 174936 app} and~\ref{tab:HD 174966 phot app} for HD\,1749436 and HD\,174966, respectively). Deformation due to rotation allowed us to provide polar and equatorial radii (see Sec.~\ref{ssec:mean density}). Some differences compared with \citet{GH11} and GH13 are expected due to the inclusion of rotation in our models, although the expectations were to find lower masses and higher volumes, while preserving mean densities. However, we got systematically lower densities. We also find higher hydrogen abundances so less evolved stars.\par

\begin{table*}
    \centering
    \caption{Characterisation of HD 174936 taking $1\sigma$ uncertainties in each observable. The second column shows the physical characterization obtained using the photometric measurements. The third column represents the values obtained when applying the large separation criteria to the photometric box subset of models. The fourth column represents the values obtained when using $\text{T}_{\text{eff}}$ and Gaia's L.}
    \begin{tabular}{c|c|c|c}
    \hline
      Parameter & Photometry &  Photometry+\Dnu &  Photometry+L$_\text{GAIA}$ \\
     \hline 
      M ($\text{M}_{\odot}$) & [1.70, 2.45] & [1.85, 2.20] & [1.80, 1.95]\\
      R$_\text{p}$ ($\text{R}_{\odot}$) & [1.53, 2.89] &  [1.96, 2.20] & [1.60, 1.80] \\
      R$\text{e}$ ($\text{R}_{\odot}$) & [1.60, 3.41] &  [2.13, 2.61] & [1.75, 2.00] \\
      L ($\text{L}_{\odot}$) & [8.54, 35.56] & [14.03, 20.79] & [10.87, 11.07] \\
      $\text{T}_{\text{eff}}$ (K) & [7800.00, 8199.98] & [7800.02, 8199.81] & [7800.07, 8198.43] \\
      logg & [3.88, 4.28] & [4.05, 4.12] & [4.17, 4.27]\\
      Xc & [0.22, 0.74] & [0.40, 0.53] & [0.53, 0.74]\\
      $\bar{\rho_{\Omega}}$ (g/cm$^3$) & [0.104, 0.599] & [0.211, 0.292] & [0.369, 0.542]\\
      \text{[Fe/H]} & [-0.4, -0.2] & [-0.4, -0.2] & [-0.4, -0.2]
    \end{tabular}
    \label{tab:HD 174936 app}
\end{table*}

\begin{table*}
    \centering
    \caption{Characterisation of HD 174966 taken $1\sigma$ uncertainties in the considered observables. Second, third and fourth columns represent the same constraints as in Table~\ref{tab:HD 174936 app} but for HD~174966. The fifth column shows the limits of the physical parameters when using the spectroscopic measurements and the sixth one when also considering \Dnu.}
    \begin{tabular}{c|c|c|c|c|c}
    \hline
      Parameter & Photometry &  Phot.+\Dnu & Phot.+L$_\text{GAIA}$ & Spec. & Spec.+\Dnu \\
     \hline 
      M ($\text{M}_{\odot}$) & [1.70, 2.30] & [1.75, 1.85] &  [1.85, 1.85] & [1.70, 1.80] & [1.75, 1.80]\\
      R$_\text{p}$ ($\text{R}_{\odot}$) & [1.63, 2.95] &  [1.69, 1.76] &  [1.85, 1.90] &  [1.58, 1.82] & [1.70,1.75]\\
      R$_\text{e}$ ($\text{R}_{\odot}$) & [1.73, 3.52] &  [1.82, 1.99] &  [2.02, 2.15] & [1.67,2.02] & [1.82, 1.95] \\
      L ($\text{L}_{\odot}$) & [7.58, 30.73] & [8.18, 11.02] & [10.74, 10.92]  & [7.34, 10.23] & [8.74, 9.33] \\
      $\text{T}_{\text{eff}}$ (K) & [7437.04, 7836.98] & [7437.92, 7836.98] & [7506.81, 7621.67] & [7505.03, 7604.95] & [7505.16, 7604.32] \\
      logg & [3.83, 4.23] & [4.19, 4.22] & [4.12, 4.16] & [4.16, 4.26] & [4.19, 4.21]  \\
      Xc & [0.20, 0.72] & [0.56, 0.61] & [0.50, 0.54] & [0.53, 0.72] & [0.56, 0.60] \\
      $\bar{\rho_{\Omega}}$ (g/cm$^3$) & [0.090, 0.489] & [0.375, 0.446] & [0.296, 0.347] & [0.345, 0.543] & [0.381, 0.438]  \\
      \text{[Fe/H]} & [-0.1, 0.0] & [-0.1, 0.0] & [0.0, 0.0] & [-0.1, 0.0] & [-0.1, 0.0] 
    \end{tabular}
    \label{tab:HD 174966 phot app}
\end{table*}

These discrepancies might come from two different sources. First, the codes used in the computation of the equilibrium models. Each of them has different prescriptions in the physics and numerical schemes used to solve the structure and evolution equations. Second, the gravity darkening: comparison between observations and modelling of rotating stars needs to consider that \teff\ and \logg\ are affected by the stellar deformation. Consequently, we did not take Gaia's luminosity and \Dnu\ simultaneously to discriminate models, since they do not show compatible restrictions (see Figure~\ref{fig:stars}).\par

Nonetheless, our conclusions are still safe. Luminosity is more affected by gravity darkening than effective temperature \citep{Paxton2019}. We can thus consider the conservative uncertainties of the photometric parameters as a secure range for the real \teff\ of our objects. We cannot consider \logg\ either. On the contrary, \Dnu\ has neither visibility nor deformation effects (because of the conservation of the \rhom-\Dnu\ relation) and it is an alternative to \logg\ (as shown by GH2017), so we can derive reliable parameters from these observables. Small uncertainties remain because of the unknown physics. Anyway, comparing Gaia's luminosities with those coming from the constraints provided by \Dnu\ might be used to get even more information about the stellar configuration, as we discuss in Sec.~\ref{sec:inclination}.

\subsection{The role of the inclination angle, $i$ }
\label{sec:inclination}

As we discussed in Sec.~\ref{ssec:mean density}, {\sc cestam}'s variables are split into a mean value over an isobar plus a perturbation. Thus, mean values of the variables corresponds to $P_2(\cos\theta) = 0$, that is, $\sin^2\theta=2/3$ \citep{PerezHernandez1999}. This means a colatitude $\theta\approx54.7356^{\circ}$, as pointed out by \citet{BarceloForteza2018}. Only the luminosity is not calculated in such a way. $L$ is the integration of the flux all over the stellar surface so it is the intrinsic stellar luminosity.\par

Then, the comparison between an observable independent of the inclination angle, such as \Dnu, and the observed luminosity will give some clue about $i$. In an HR diagram, a star with $i<55^{\circ}$ would show a higher projected luminosity than the intrinsic one. And, on the contrary, when $i>55^{\circ}$, the projected luminosity would be lower \citep[see][for examples of both cases]{Paxton2019}.\par

Our sample stars show both situations. HD\,174936 shows a lower Gaia luminosity than \Dnu\ so $i>55^{\circ}$. That means a bounded rotational velocity between $\text{v}_\text{rots} = [169.7, 207.2]$~km/s. On the other hand HD\,174966 is the opposite, with a higher Gaia's luminosity and $i<55^{\circ}$. GH13 estimated an inclination angle of $i=[45^{\circ}, 70^{\circ}]$, based on a LPVs (Line Profile Variations) analysis, with the most probable value at $i=62.5^{\circ}$. With all these considerations, the rotational velocity of this star seems to be in the range of $\text{v}_\text{rots} = [135, 153]$~km/s, corresponding to $i=[45^{\circ}, 55^{\circ}]$. In any case, this angle must not be very far from $55^{\circ}$, since Gaia's and \Dnu\ luminosities slightly overlap for the photometric \teff.\par

A more detailed analysis could give a more precise value of $i$. This is not the scope of this research but it will be investigated in a forthcoming work.

\section{Conclusions}\label{sec:conclusions}

We have conducted a study of the impact of rotation on the relation between the predicted large separation and mean density in the low order regime. This work updates the study performed by \citet{Suarez2014} with non-rotating models. Here we have updated that work by assessing in detail the effect of the moderate rotation on the \Dnu -\rhom\ relation. We have computed a grid of rotating asteroseismic models representative of intermediate-mass stars. \par

We have first confirmed the consistency of our modelling when computing the mean density of the models on a deformed surface (ellipsoid). The fits obtained are closer to the empirical ones than those obtained with \rhom\ from spherically-symmetric models. Although differences do not change the overall behaviour, they become important when those fits are employed to characterise individual stars. For example, mean densities are clearly different when not considering a spheroid volume.\par 

The linear relation for the rotating models was found in line with the previous predictions \citep{Suarez2014}, as well as with empirical results \citep{GH2015, GH2017}. Like in those works, we have found a dispersion in this relation, explained mainly by the stellar mass. The detailed comparison by mass ranges (buckets), yields a systematic variation of the coefficients all along the whole mass range. The majority of the individual fits remain within those found by the aforementioned works, favouring the matching for highest masses. Additionally, most of the individual fits for masses larger than $1.8\,\mathrm{M}_\odot$ are closer to the empirical one (GH17) except for the very highest masses close to $3\,\mathrm{M}_\odot$, making it even more reliable for \dss.

We carried out an extra test to characterise the parameter $\varepsilon$ from a fit to the frequencies following the classical asymptotic relation: $\nu = \Delta\nu(n+\varepsilon)$, also used in \citet{Bedding2020}. \textbf{In contrast to their findings}, we could not establish a clear relation between this parameter and the evolutionary state of the model. This is maybe due to the dependency on rotation but additional work must be done in this sense to clarify these conclusions.

Following the methodology used in \citet{GH2013}, we applied the new fit obtained for rotating stars to constrain global parameters of the $\delta$ Scuti stars HD\, 174936 and HD\,174966, with the objective of comparing the results with previous works in which rotation was not taken into account. The models constrained this way were then confronted with the observed temperatures and surface gravities obtained from spectroscopy and photometry, including luminosities from Gaia's DR2 data. For both stars we predict systematic lower densities as well as higher hydrogen abundances than for the non-rotating case. This would imply that both stars are less evolved than has been characterised in GH09 and GH13. \par

Additionally, the discrepancies found in the luminosities derived from \Dnu\ and those observed by Gaia allow us to discuss the effects of gravity darkening in our study. In particular, for HD\,174936, the analysis of limb darkening allowed us to determine a lower bound for the angle of inclination ($i> 55~\deg$), and thereby the surface rotational velocity is bounded to $v_\mathrm{rots} = [169, 207]~\mathrm{km/s}$. On the other hand, for HD\,174966 we found an upper bound for the angle of inclination ($i<55~\deg$), which allowed us to constrain the surface rotational velocity to $v_{\mathrm{rots}} = [135, 153]~\mathrm{km/s}$. This implies an angle of inclination of $i=[45, 55]~\deg$, which a smaller value than GH13's predictions (although still compatible with that work's results within the $1\sigma$ uncertainty in $i$).\par

We have thus (1) confirmed the robustness and reliability of the empirical \Dnu-\rhom\ relation given by GH17, providing new limits to the relation (Eq.~\ref{eq:fit}) that encompass any possible dependency on mass; and (2) found a reliable methodology using 1D-evolutionary rotating models and perturbative theory for characterising moderately-fast rotating stars. This methodology allows us to constrain the angle of inclination of the star and hence the actual surface rotational velocity. Yet this constraint is still small, and additional information on $i$ is required. We plan to deepen on this by studying objects for which both high-resolution spectroscopy and interferometry measurements are available (e.g. the stars Altair or Rasalhague).\par

Furthermore, we want to investigate how our method can be used in the construction of 2D rotating models, e.g. with the ESTER code, as it may provide sufficiently accurate constraints to get the \emph{seed}-1D model for the 2D computations.\par

\section*{Acknowledgements}
The authors want to strongly thank Dr. Daniel R. Reese for his very professional and useful report that helped us to significantly improve the paper.

As well, the authors acknowledge funding support from Spanish public funds (including FEDER funds) for research under projects ESP2017-87676-C5-2-R ESP2017-87676-C5-5-R, and PID2019-107061GB-C64. JERM also acknowledges financial support from the State Agency for Research of the Spanish MCIU through the ``Center of Excellence Severo Ochoa" award to the Instituto de Astrof\'{\i}sica de Andaluc\'{\i}a-CSIC (SEV-2017-0709). AGH also acknowledges support from ``FEDER/Junta de Andaluc\'{\i}a-Consejer\'{\i}a de Econom\'{\i}a y Conocimiento'' under project E-FQM-041-UGR18 by Universidad de Granada. JCS also acknowledges support from project RYC-2012-09913 under the `Ram\'on y Cajal' program of the Spanish Ministry of Science and Education. This work has made use of data from the European Space Agency (ESA) mission {\it Gaia} (\url{https://www.cosmos.esa.int/gaia}), processed by the {\it Gaia} Data Processing and Analysis Consortium (DPAC, \url{https://www.cosmos.esa.int/web/gaia/dpac/consortium}). Funding for the DPAC has been provided by national institutions, in particular the institutions participating in the {\it Gaia} Multilateral Agreement.

This is a pre-copyedited, author-produced PDF of an article accepted for publication in Monthly Notices of the Royal Astronomical Society following peer review. The version of record \cite{2020MNRAS.tmp.2504R} is available online at: \url{https://academic.oup.com/mnras/article-abstract/498/2/1700/5897367?redirectedFrom=fulltext}

\section*{Data Availability}
The data underlying this article will be shared on reasonable request to the corresponding author.




\bibliographystyle{mnras}
\bibliography{bibliography.bib} 











\bsp	
\label{lastpage}
\end{document}